\documentclass[runningheads]{llncs}

\usepackage[utf8]{inputenc}
\usepackage{graphicx,color}
\usepackage{tikz}
\usepackage{authblk}
\usepackage{amsmath,mathrsfs,amsfonts,latexsym,amssymb,verbatim}
\usepackage{algpseudocode}
\usepackage{latexsym}
\usepackage{enumerate}
\usepackage{mathtools}
\usepackage{hyperref}
\usepackage[export]{adjustbox}
\usepackage{microtype}
\usepackage{relsize}
\usepackage[normalem]{ulem}
\usepackage{mdwlist} 
\usepackage{caption}
\newcommand{\VL}[1]{}

\definecolor{brightcerulean}{RGB}{0,103,147}
\definecolor{brightother}{RGB}{167,123,0}

\newcommand{\ov}[1]{\overline{#1}}
\newcommand{\stab}[1]{\mathcal S(#1)}

\def\J{\mathcal J}
\def\S{\mathcal S}
\def\Z{\mathcal Z}
\def\X{\mathcal X}
\def\T{\mathcal T}
\def\L{\mathcal L}
\def\F{\mathcal F}

\def\Y{\mathcal Y}
\def\G{\mathcal G}
\def\A{\mathcal A}

\newcommand{\gquotient}[2]{#1{/#2}}

\usepackage{ifthen}
\newcommand{\ifnv}[2]{\ifthenelse{\equal{#1}{}}{}{#2}}
\newcommand\sett[3][]{\left\{\left.#2\ifnv{#1}{\in #1}\vphantom{#3}\right|#3\right\}}
\newcommand\restr[1]{_{\left|#1\right.}}
\newcommand\dfn[1]{\emph{#1}}
\newcommand\card[1]{\left|#1\right|}
\newcommand\subfini{\subset}

\newcommand{\setR}{\mathbb{R}}
\newcommand{\setZ}{\mathbb{Z}}
\newcommand{\setF}{\mathbb{F}}

\newcommand\cyl[2][]{\left[#2\right]_{#1}}
\newcommand\defeq{:=}
\newcommand{\ie}{\textit{i.e.}\ }
\newcommand\conc{} 

\begin{document}
\title{Graph subshifts}
\titlerunning{Graph subshifts}

\author{Pablo Arrighi\inst1\inst2 \and Amélia Durbec\inst3 \and Pierre Guillon\inst4}
\institute{Université Paris-Saclay, Inria, CNRS, LMF, 91190 Gif-sur-Yvette, France\and IXXI, Lyon, France \and Université Paris-Saclay, CNRS, LISN, 91190 Gif-sur-Yvette, France\and Aix-Marseille Université, CNRS, I2M, Marseille, France
}
\authorrunning{P. Arrighi, A. Durbec, P. Guillon}




\maketitle


\abstract{We propose a definition of graph subshifts of finite type that can be seen as extending both the notions of subshifts of finite type from classical symbolic dynamics and finitely presented groups from combinatorial group theory. These are sets of graphs that are defined by forbidding finitely many local patterns. In this paper, we focus on the question whether such local conditions can enforce a specific support graph, and thus relate the model to classical symbolic dynamics. We prove that the subshifts that contain only infinite graphs are either aperiodic, or feature no residual finiteness of their period group, yielding non-trivial examples as well as two natural undecidability theorems.}

\section{Introduction}
Subshifts of finite type are well studied objects in symbolic dynamics \cite{LindMarcus} and ergodic theory. Given an alphabet $\Sigma$, a subshift of finite type (SFT) on $\setZ^d$ is a set of configurations $\Y \subseteq \Sigma^{\setZ^d}$ that do not contain a given finite set of forbidden patterns $\F \subseteq \Sigma^{\{-n,n\}^d}$. In spite of their relatively benign and local definition, SFT have proven to have complex global behaviours. One example of such results is the existence of aperiodic SFTs in dimension $2$, but not in dimensional $1$. Even some natural problems such as determining whether an SFT contains a periodic configuration at all, are proven to be undecidable \cite{Berger}. This local-to-global complexity is in fact shared by multiple dynamical systems and most notably with Cellular Automata (CA) \cite{Langton}.
However, the two models are deeply connected as CA of dimension $d$ can be seen as subshifts of dimension $d+1$.
This connection as been used to prove multiple theorems on different subclasses of CA \cite{Kari}.  This result is also one of the original motivations of this paper: Graph subshifts ought to be a means to study a generalization of CA called Causal Graph Dynamics (CGD) \cite{DurbecIMRCGD}. CGD extend CA in two complementary ways: first because they are defined on arbitrary graphs of bounded degree $m$, and second because they allow for the graph itself to evolve, according to a local, shift-invariant rule. Graph subshifts aim to encompass CGD spacetime diagrams as a subclass.

Subshifts have already been generalized to configurations over 
finitely generated groups.
However this is not enough to simulate CGD as all configurations share the same support graph. In this paper, we provide a formalism which relaxes this constraint, allowing the support of the graphs itself to also be prescribed by set of forbidden patterns. We introduce and formalize the notion of graph subshifts, i.e. sets of graphs that are defined by forbidding local patterns. 

The natural question to ask, then, is whether we may enforce, by means of such local constraints only, that the set of graphs be of a particular shape. In particular, notice that a graph subshift may contain a finite or an infinite number of graphs (up to shift), and that the graphs themselves may be of finite or of infinite size. In this first paper we focus on the question whether there exist subshifts whose graphs are all of infinite size. The question is non-trivial; for instance we prove that the problem whether a given set of forbidden patterns induces a graph subshift without finite configuration is undecidable. The problem whether it uniquely fixes the support graph is also shown undecidable. Still, we prove that the graph subshifts that contain only infinite graphs are either aperiodic, or feature no residual finiteness of their period group, yielding us with non-trivial examples and establishing connections with different areas of mathematics: periodicity from symbolic dynamics; residually finite groups from combinatorial group theory \cite{BaumslagSolitar}\cite{LyndonSchupp}; and graph covers as used in graph theory and distributed computing.

All proofs are in the Pre-print version of this paper. A very exploratory version of those results has been presented in the (non-proceedings track of) workshop Automata 2021.

\section{Graphs}\label{sec:graphs}
The notions of graphs and of pointed graphs modulo, together with their operations, are intuitive enough as summarized here, for rigorous definitions see \cite{ArrighiCayleyNesme}. Let $\pi$ be the finite set of ports, $\Pi=\pi^2$, and $V$ some universe of names.\\
\noindent {\em Graphs.} The graphs denoted by letters $G, H\ldots$ are the usual, connected, undirected, possibly infinite, bounded-degree labelled graphs, but with a few added twists:
\begin{itemize}
\item Vertices are connected through their ports. An edge is an unordered pair $\{x : a, y : b\}$ of $V\times\pi$, or a singleton $\{x : a\}$ standing for a self-loop, where $x, y$ are vertices and $a,b\in \pi$ are ports.
Each port is used at most once per node: if $e,e'\in E(G)$ intersect, they must be equal.
As a consequence the degree of the graph is bounded by $\card\pi$.
\item Each vertex $x$ is given a label $\sigma(x)$, taken within a finite set $\Sigma$, also referred to as an internal state. Sometimes the function $\sigma$ is denoted $\sigma_G$ to emphasize that it belongs to the description of $G$.
\end{itemize} 
The set of all (finite and infinite) graphs having ports $\pi$, vertex labels $\Sigma$ is denoted $\G_{\Sigma,\pi}$, or simply $\G$.
The \dfn{support} of graph $X$ is simply the structure of vertices linked by edges with ports, forgetting the labeling $\sigma$, or equivalently the graph in which the labeling is changed to the monochromatic one.

\begin{figure}
\captionsetup{width=0.83\textwidth}
\begin{center}
\includegraphics[scale=.7,clip=true,trim=0cm 0.22cm 0cm 0.1cm]{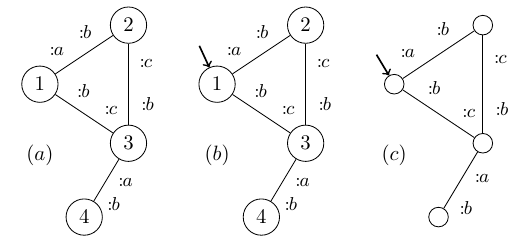}
\end{center}
\caption{\label{fig:graphs} (a) A graph $G$. (b) A pointed graph $(G,1)$. (c) A pointed graph modulo $X$.}
\end{figure}


\noindent {\em Pointed graphs modulo.} The goal of this paragraph is to define a set of graphs that is a compact space. Indeed, consider a graph $G$, a vertex $p\in V(G)$, and $D=G^r_p$ the disk of radius $r$ around $p$, \ie the subgraph of $G$ induced by the $r$-neighbors of $x$ in $G$. Now forgetting about $G$, this $D$ is one amongst the many possible disks of radius $r$ that can appear in $\G$.
In order for the set of graphs to form a compact space, one requirement is that the set of disks of radius $r$ be finite. Morally this is the case, since the degree is bounded by $|\pi|$, but vertex names $x,y,z$ jeopardize this; we must get rid of them. Thus, 
\begin{itemize}
\item Graphs are endowed with a privileged pointed vertex $p$ playing the role of an origin (in particular, there is no empty graph).
\item These pointed graphs $(G,p)$ are then considered modulo isomorphism. Notice that isomorphism rename the pointer in the same way as they rename the vertex upon which it points, i.e. $R:(G,p)\mapsto(RG,R(p))$. Thus the relative position of the other vertices w.r.t to the pointer is unchanged. 
\end{itemize} 
Pointed graphs modulo are denoted $X, Y\ldots$. The set of all pointed (finite and infinite) graphs modulo is denoted $\X_{\Sigma,\pi}$, or simply $\X$. Given a graph non-modulo $G$, and a vertex $p\in V(G)$, we write $X((G,p))$ for the corresponding pointed graph modulo. Additionally the {set of disks of radius $r$} with states $\Sigma$ and ports $\pi$ is written $\X_{\Sigma,\pi}^r$, or simply ${\X}^r$.
\\
\noindent {\em Paths and vertices.}
Over graphs modulo isomorphism without pointer, vertices no longer have a unique identifier, which makes designating a vertex a daunting task.
Fortunately, our graphs are pointed, so that any vertex $\hat{u}$ can always be designated by some sequence of ports $u$ that leads to it, starting from the origin.
Given a pointed graph modulo $X$, its set of finite paths, starting from the origin, forms a subset $\L(X)$ of $\Pi^*$ with $\Pi=\pi^2$.
Whenever two paths $u$ and $v$ lead to the same vertex, we write $u\equiv_X v$.
Thus vertex $\hat{u}$ is really the $\equiv_X$-equivalence class of $u$, hence the hat notation.
The set of these equivalence classes, a.k.a vertices, is denoted $V(X)$.
Notice that, starting from origin, following path $v=(ab)\cdots(cd)$ and then path $\ov{v}=(dc)\cdots(ba)$, leads back to the vertex at the origin, namely $\hat{\varepsilon}$, where $\varepsilon$ denotes the empty path.

  \begin{remark}\label{r:equiv} Notice that two paths $u,u'\in \L(X)$ designate the same vertex $\hat u=\widehat{u'}$ if and only if for every $v\in \L(X_u)$, $\widehat{uv}=\widehat{u'v}$ (in particular, $v\in \L(X_{u'})$).
    \end{remark}

\noindent {\em Back to graphs non-modulo.}
Sometimes we still need to manipulate usual graphs, where vertices do have names. For this purpose we use a canonical naming function $X\in\X\mapsto G(X)\in \G$, which names each vertex of $X$ by the set of paths that lead to it, starting from $\varepsilon$. This $G(X)$ is referred to as `the associated graph'. One has $X(G(X),\hat{\varepsilon})=X$.
\\
\noindent {\em Shift.}\label{def:shift}
Let $X\in {\X}$ be a pointed graph modulo.
Consider a path $u\in \L(X)$. Then $X_u$ is $X((G(X),\hat{u}))$. The pointed graph modulo $X_u$ is referred to as {$X$ \emph{shifted by} $u$}.

From now on we refer to $X\in{\cal X}$ as just `a graph', as it is clear from the notation that it is in fact a pointer graph modulo. 

\section{Subshifts}\label{sec:subshifts}

Given a pointed graph modulo $X$, a word in $\Pi^*$ designates a vertex, and a language $L\subseteq {\Pi}^*$ designates a set of vertices.
If, moreover, $L$ is stable under taking the prefix, then the designated vertices induce a connected pointed subgraph of $X$ which we refer to as `cut'.
Given a set $\Z$ of graphs, one may be interested in selecting those whose cut according to $L$ equals some graph $Z$, which we refer to as `cylinder'. 

\begin{definition}[Cuts and cylinders]
Consider $L\subseteq {\Pi}^*$ a prefix-stable language, \ie such that $u\conc v\in L$ implies $u\in L$.
The \dfn{$L$-cut} of a graph $X\in {\cal X}$ is the subgraph induced by the vertices $\sett{\hat{u}}{u\in L\cap \L(X)}$.
It is denoted $X\restr L$. 
Consider ${\Z}\subseteq{\cal X}$, we write $\Z\restr{ L}$ for $\sett{X\restr L}{X\in\Z}$.
Consider $Z\in\Z\restr L$, the \dfn{cylinder} of $Z$ within $\Z$ is $\sett{X\in\Z}{X\restr L=Z}$.
It is denoted $\cyl[\Z]Z$.
\end{definition}

\begin{definition}[Subshift]
Let $\F$ be a set of tuples $(F,L)$, where each $F$ is a finite graph, and each $L$ a {finite} prefix-stable language.
The \dfn{subshift} forbidding $\F$ is 
\[\Z=\sett[\X_{\Sigma,\pi}]X{\forall v\in \L(X),\forall (F,L)\in\F,(X_v)\restr{L}\ne F}.\]
{It is of finite type if $\F$ can be chosen finite.}
\end{definition}

\begin{remark}
{In the pairs $(F,L)$, $L$ cannot always be assumed to be the language labelling the paths of $F$. For example, requiring every vertex to have an edge with port $a\in\pi$ is a very natural finite-type condition, but this SFT is not of the form $\sett[\X_{\Sigma,\pi}]X{\forall v\in \L(X),\forall F\in\F,(X_v)\restr{ L_F}\ne F}$ as the $ L_F$ do not contain $a$ and hence cannot tell that it is absent.\\
Yet, this SFT can be defined through our definition by 
$$\sett{(F,\{\varepsilon,ap\})}{p\in\pi, F\text{ 1-node graph}}.$$
  The same remark holds for the dual definitions, \ie through allowed patterns: think of the SFT consisting of those graphs having no port $a\in\pi$.}
\end{remark}

\begin{lemma}\label{l:sft}
The following are equivalent.
	\begin{enumerate}
		\item\label{i:forbs} $\Z$ is a subshift of finite type.
		\item\label{i:forb} $\Z$ is the subshift forbidding some $\F'\times\{M\}$ where $\F'$ is a finite set of finite graphs and $M$ a single finite prefix-stable language.
		\item\label{i:allos} $\Z$ is the set of graphs allowing some finite set $\A$ of tuples $(A,L)$, where $A$ is a finite graph and $L$ a finite prefix-stable language---in the sense that:
		\[\Z=\sett[\X_{\Sigma,\pi}]X{\forall v\in \L(X),\exists (A,L)\in\A,(X_v)\restr{L}=A}.\]
		\item\label{i:allo} $\Z$ is the set of graphs allowing $\A'\times\{M\}$ where $\A'$ is a finite set of finite graphs and $M$ a single finite prefix-stable language.
	\end{enumerate}
\end{lemma}

\begin{definition}[Defining window] 
The prefix-stable language $M$ is the same in Conditions~\ref{i:forb} and~\ref{i:allo} of Lemma~\ref{l:sft}.
It is called a \dfn{defining window} for $\Z$.
\end{definition}

\begin{remark}
{The duality between forbidding and allowing no longer holds in general when the subshifts are not of finite type. For example we can think about the set of graphs authorizing exactly those finite graphs which have exactly one red vertex.	
This set however is not closed topologically (because the red vertex can in some sense vanish to infinity; cf. \cite{ArrighiCayleyNesme} as summarized by the Appendix) and hence it is not a subshift.
Its closure is the so-called \dfn{one-dimensional sunny-side-up} subshift, as defined by forbidding those finite graphs having two red vertices.}
\end{remark}

%

\begin{remark}
	Given any subshift $\Z$, one can define the subshift of its infinite graphs, by additionnally forbidding $(F,L\pi^2)$ for every prefix-stable language $L\in{\Pi}^*$ and every graph $F$ such that $F\restr L=F$. Indeed this condition exactly means that $L$ was already describing all possible vertices in $F$, and longer paths do not add any.
	Nevertheless, this subshift is very rarely of finite type, as we will see in the following section.
\end{remark}
On the contrary, all SFTs presented so far contain finite graphs.
It is not obvious a priori whether some SFTs contain only finite graphs, or even whether they are {homogeneous}, in the sense that all the graphs share a common support.
We will discuss these properties in the next sections.

\section{Cayley SFTs}
The equivalent definitions for SFTs all capture the concept of local constraint, and in the following sections, we often define our SFTs by describing some conditions that graphs must locally satisfy.
Among these conditions, \dfn{nearest-neighbor constraints} are those defined with a single edge between two vertices, that is $L\subseteq\{\varepsilon\}\sqcup\Pi$.
\begin{example}
  A \dfn{directed $2$-regular SFT} is an SFT with ports $\pi=\{a,a',b,b'\}$ obtained by imposing nearest-neihbor constraints:
  \begin{enumerate}
    \item enforcing that edges be of the form $pp'$;
    \item enforcing that all ports be occupied.
  \end{enumerate}
  It contains, amongst many others, the Cayley graph of the free group over two generators (a tree), as well as those of all of its quotients.
\end{example}



\begin{figure}
\captionsetup{width=0.83\textwidth}
\begin{center}
\begin{tabular}{c  c}
 \includegraphics[scale=1]{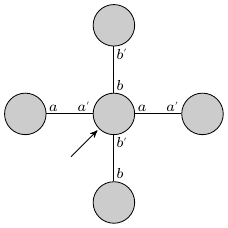}  &  \includegraphics[scale=1]{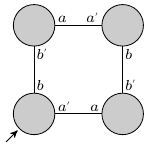} \\
 $ L =\{\varepsilon, aa', a'a, bb', b'b\}$ & $ L =\{\varepsilon, aa', bb', aa'bb'\}$ 
\end{tabular}
\end{center}
\caption{\label{fig:monochromatic}  \emph{Locally grid-like SFT corresponding to the Cayley graph of $\langle a,b|aba^{-1}b^{-1} \rangle$}. The SFT is defined by forbidding all but these patterns, on their corresponding languages.}  
\end{figure}

\begin{example}[Locally grid-like SFT]
  We can define a graph SFT with ports $\pi=\{a,a',b,b'\}$ corresponding to the monochromatic fullshift on $\langle a,b|aba^{-1}b^{-1}=\varepsilon \rangle$, by imposing three families of constraints: \begin{enumerate}
  \item enforcing that edges be of the form $pp'$;
  \item enforcing that all ports be occupied;
  \item enforcing that the subgraph induced by $\{\varepsilon, aa', bb', aa'bb'\}$ obeys the relation \ie form a square as in Fig. \ref{fig:monochromatic}).
  \end{enumerate}
  This graph SFT obviously contains the Cayley graph of $\langle a,b|aba^{-1}b^{-1}=\varepsilon \rangle$, but also the Cayley graphs of all quotient groups {(excluding really small groups such as the trivial group)}.
\end{example}

\setlength{\tabcolsep}{1cm} 


%

A similar construction can be performed for any finitely presented group, and via a more precise third constraint, get a cleaner statement (with all possible quotients).

\begin{definition}[Cayley SFT]
	Consider a finitely presented group $\Gamma=\langle\J|\mathcal R\rangle$, that is, a group generated by the abstract finite set $\J$, where $\mathcal R$ represents a set of relations, that is words over $\J\sqcup\J^{-1}$ that represent the identity element (and from which all such words can be derived).
	For example $\setZ^2=\langle a,b|aba^{-1}b^{-1}\rangle$ (it will always be implicitly given by this canonical presentation) and $\setF_2=\langle a,b|\rangle$.
	
	For every alphabet $\Sigma$, there is a canonical way to associate a \dfn{Cayley SFT} $\Z^\Gamma_\Sigma$ to the group $\Gamma$ (in fact, to the pair $(\Gamma,\J)$, which is a \dfn{marked group} when $\J$ is symmetric and does not contain the identity $\varepsilon$).
	Define $\pi=\J\sqcup\J^{-1}$ and the following constraints:
	\begin{enumerate}
		\item each edge involves two inverse ports;
		\item every vertex must have all possible ports;
		\item for every word $u\in\mathcal R$ and $L$ its language of prefixes, one forbids the pair $(Y,L)$ for every graph $Y$ supported by $L$ in which path $u$ is not a cycle back to the origin.
	\end{enumerate}

        Let $\Gamma$ be a finitely presented group. A SFT over alphabet $\Sigma$ is defined by \dfn{$\Gamma$-NN constraints} if they include precisely the forbidden patterns defining $\Z^\Gamma_\Sigma$ and some additional nearest-neighbor constraints.
        Note that if $\Gamma=\setF_k$, then $\setF_k$-NN constraints are simply constraints on the edges, provided the correct regularity of the graph.
  \end{definition}
  \begin{remark}
The Cayley SFT obviously contains the Cayley graph of $\Gamma$, but also all Cayley graphs of its quotient groups, as will be seen in the next section.
For example, $\Z^{\setZ^2}_\Sigma$ contains the infinite grid as well as tori and cylinders, and $\Z^{\setF_2}_\Sigma$ contains Cayley graphs of all groups with two generators.

Any classical nearest-neighbor SFT defined over some Cayley graph $\Gamma$ is included in the corresponding graph SFT defined by $\Gamma$-NN constraints. 

To define a subshift whose unique support is the Cayley graph of $\Gamma$, one generally needs to add many constraints imposing that paths that are not labeled by words representing the identity element be not cycles. This usually yields a subshift of infinite type.
\end{remark}

\begin{figure}
\captionsetup{width=0.83\textwidth}
	\begin{center}
		\includegraphics[scale=.6, clip=true, trim=0cm 0.5cm 0cm 0.5cm]{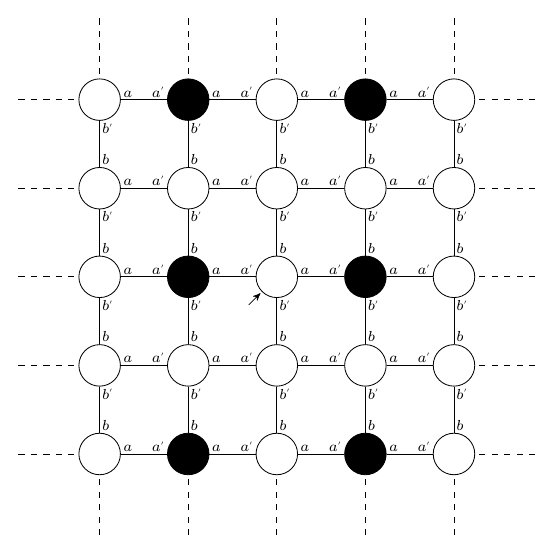}
	\end{center}
	\caption{\label{fig:hardsquaregridconfig}  \emph{Configuration of the hard-square model that is present in $\setZ^2$}. The colours represent the $0/1$ internal states of the vertices.}
\end{figure}
\begin{figure}
\captionsetup{width=0.83\textwidth}
 \includegraphics[width=0.83\textwidth, center,scale=1]{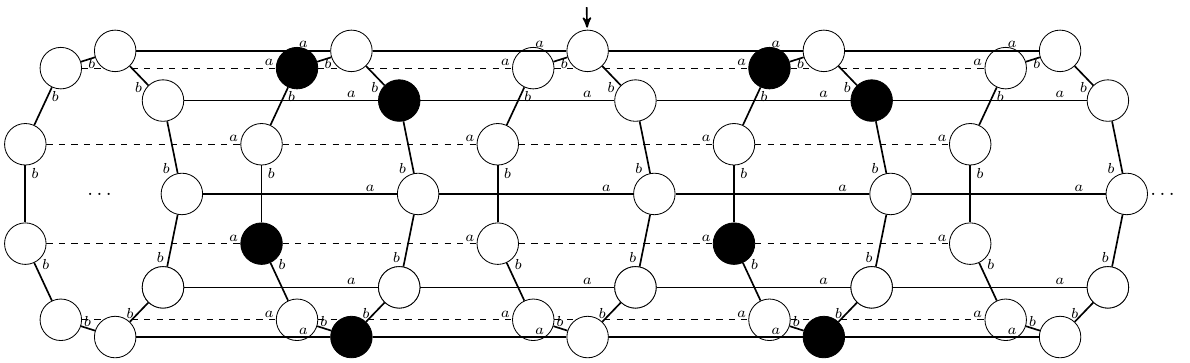}
  \caption{\label{fig:hardsquarecylconfig}  \emph{Configuration of the hard-square model whose support graph is not $\setZ^2$}. In order to preserve clarity, only half of the ports are written and some edges have been dashed.}
\end{figure}

\begin{example}[Hard-square model]
The hard-square model SFT is defined by $\setZ^2$-NN constraints: $\Z \subseteq \X_{\{0,1\},\{a,a',b,b'\}}$ such that for every configuration $X$ and edge $\{x: a,y: b\} \in E(X)$, one has that $\sigma_X(x) = 0$ or $\sigma_X(y) = 0$.
This can be expressed as an SFT by forbidding the $0/1$-coloured versions of Fig. \ref{fig:monochromatic} that have two adjacent $1$-coloured vertices.
This is a direct generalization of a well-known SFT over $\setZ^2$.
As a graph SFT, it not only contains all the configurations that are already present on $\setZ^2$ such as in Fig. \ref{fig:hardsquaregridconfig}, but also new infinite configurations such as the cylinder of Fig. \ref{fig:hardsquarecylconfig}, and even finite configuration such as tori only containing $0$.
\end{example}


Besides finitely presented groups, the notion of quotient of a group can be extended to periodic graphs, which is the subject of the following section.

\section{Periodicity and finiteness}
\subsection{Quotient and Periods}


\begin{definition}[Period]
  Any word $u\in \L(X)$ such that $X=X_u$ is called a \dfn{path-period}. 
  The corresponding vertex $\hat u$ is called a \dfn{period}.
The \dfn{stabilizer} of $X$ is its set of periods; it is denoted by $\stab{X}$.
$X$ is called \dfn{weakly periodic} if $\stab X\supsetneq\{\hat\varepsilon\}$, \dfn{strongly nonperiodic} if $\stab X=\{\hat\varepsilon\}$.
\end{definition}

Notice that if $X$ is weakly periodic, then $\stab X$ is infinite or every period $\hat u$ is \dfn{torsion}, \ie $\widehat{u^n}=\hat \varepsilon$ for some $n\ge2$.


Notice that monochromatic Cayley graphs are exactly the graphs whose all vertices are periods.
As we have seen in our Cayley SFT examples, periodic graphs can typically be ``folded into smaller graphs'' that also belong to the graph SFT. The following formalizes the idea of a graph folding into another one.

\begin{definition}[Homomorphism]\label{d:morph} A \dfn{homomorphism} from graph $X\in\X$ to graph $X'\in\X$ is 
a function $\varphi$ from $V(X)$ to $V(X')$ such that for all $u \in \L(X)$,
	\[\varphi(\hat{u})=\hat{u}\quad and\quad\sigma_{X}(\hat u)=\sigma_{X'}(\hat u).\]
          Note the first equality implies $\L(X)\subseteq \L(X')$.
          When $\L(X)=\L(X')$ (note that this implies that $\varphi$ is surjective and locally injective), we then say that $X$ \dfn{covers} $X'$ and that $\varphi$ is a \dfn{covering}. 
\end{definition}

\begin{remark}
  A homomorphism sends cycles to cycles and does not change the set of path-periods (i.e. if $u'$ is such that $\hat u'=\phi(\hat u)$, then $X=X_u\iff X'=X'_{u'}$).
  Hence if $\hat{u}$ is a period of $X$, $\varphi(\hat{u})$ is a period of $\varphi(X)$.
\end{remark}

\begin{definition}[Quotient graph]
          Consider $X\in {\cal X}$ and a stabilizer subgroup $H$.
 We can define a quotient graph $\gquotient{X}{H}$ to be $X((G,p))$ where:
\begin{itemize}
	\item $V(G) = \sett{H\conc\hat{u}}{\hat{u} \in V(X)}$, \ie each vertex is labelled by a set $H\conc\hat{u}=\sett{Hv}{v\equiv_Xu}$.
	\item $p=H$, \ie the pointer is the vertex labeled by $H$.
	\item $E(G) = \sett{(H\conc\hat{u}:a, H\conc\widehat{u\conc ab}:b)}{u\conc ab\in \L(X)}$.
	\item $\sigma_{G}(H\conc\hat{u}) = \sigma_{{X}}(\hat u)$.
\end{itemize}
\end{definition}

By stabilizer subgroup, we mean a subgroup of the stabilizer group. As the elements of a stabilizer subgroup cannot be distinguished modulo isomorphism, it follows that a stabilizer subgroup induces a natural covering: 

\begin{proposition}
  \label{p:quotient}
For every graph $X\in\X$ and every stabilizer subgroup $H$ of $X$,
  $\gquotient{X}{H}$ is a well-defined graph, and $\phi:\hat u\mapsto H\conc\hat u$ is a covering from $V(X)$ to $V(X/H)$.
\end{proposition}

On top of being able to "fold" a graph, we can do so while locally preserving it, given that a certain condition on the stabilizer subgroup is met:

\begin{definition}[Separated]
  If $U,V\subset\Pi^*$, then we denote $U\conc V\defeq\sett{u\conc v}{u\in U,v\in V}$, and $\ov U\defeq\sett{\ov u}{u\in U}$.
  If $M\subseteq\Pi^*$ is a prefix-stable language, we say that a set $H$ of vertices is \dfn{$M$-separated} in graph $X$ if for every $\hat u\in V(X)$, there at most one $\hat v\in H$ such that there exists $w\in M\ov M$ with $\widehat{uw}=\hat v$.
\end{definition}

\begin{proposition}\label{p:rseparate}
  Let $X$ be a graph, $M\subseteq\Pi^*$ a prefix-stable language, and $H\le\stab{X}$ a stabilizer subgroup which is {$M$}-separated in $X$.
  Then the covering preserves the language of support $M$ in the sense that for all $u \in \L(X)$, $(X_u)\restr{M}=(\gquotient{X}{H}_u)\restr{M}$ (modulo isomorphism).\\
	In particular, if $X$ belongs to some SFT $\X$ with defining window $M$, then $\gquotient{X}{H}$ also belongs to $\X$.
      \end{proposition}

\subsection{Strong periodicity}
\begin{definition}[Density, strong periodicity]
  Consider a graph $X$ and a finite prefix-stable language $M\subfini\Pi^*$.
  We say that a language $L\in \Pi^*$ is \dfn{$M$-dense} in $X$ if for all $\hat{u}\in V(X)$, $u\conc M\cap L\ne\emptyset$.
  $L$ is said dense in $X$ if such a $M$ exists.
  A graph $X$ is said to be \dfn{strongly periodic} if its stabilizer $\stab{X}$ is dense in $X$.
  Equivalently, there are only finitely many distinct (modulo isomorphisms) graphs $X_u$, for $u\in \L(X)$.
      
\end{definition}
Note that finite graphs are strongly periodic, but may not always be weakly periodic.
Apart from this degenerate case, strongly periodic graphs are weakly periodic.

Notice also that strong periodicity is almost equivalent to being a Cayley graph, up to replacement of each vertex by a fixed finite graph (with possible edges between them). This replacement (factor map) corresponds somehow to the application of a causal graph dynamics \cite{ArrighiCayleyNesme} to the Cayley graph.

\begin{lemma}\label{l:finquot}
A graph $X$ is strongly periodic if and only if $\gquotient XH$ is finite for some stabilizer subgroup $H$.
\end{lemma}
{
The following result establishes a connection between the SFT that admit finite graphs, and the well-established notion (see for instance \cite[Chap.~2]{cagrp}) of residual finiteness from group theory: 
\begin{definition}[Residually finite group]
A group ${\S}$ is residually finite if and only if for any non-identity element $\hat{u}\in {\S}$, there exists a normal subgroup $H\trianglelefteq{\S}$ of finite index such that $\hat{u}\notin H$.
\end{definition}
This notion is very robust. In particular, the property extends to finite sets of non-identity elements $I$, \ie there exists a normal subgroup $H\trianglelefteq{\S}$ of finite index such that $H\cap I=\emptyset$, simply because any finite intersection of finite-index subgroups is a finite-index subgroup.

Let us relate residual finiteness to quotient graphs, at the intuitive level first. Say $X$ belongs to some SFT ${\cal Y}$ and let $\hat{u}_1, \ldots, \hat{u}_n$ be distinct elements of $V(X)$.
The above property gives a subgroup $H$ of its stabilizer, and thus $\gquotient{X}{H}$ is a possible quotient graph; one which maintains that $\hat{u}_1, \ldots \hat{u}_n$ are distinct.
The correct choice of $\hat{u}_1, \ldots, \hat{u}_n$ will help us make sure that $\gquotient{X}{H}$ still belongs to SFT ${\cal Y}$.

\begin{theorem}\label{t:perfin}
  Let $\Y$ be a SFT. $\Y$ contains a finite graph if and only if $\Y$ contains a strongly periodic graph $X$ such that the stabilizer $\stab{X}$ is residually finite.
\end{theorem}

Thus there are two complementary ways in which a nonempty SFT may have only infinite configurations: either by preventing strongly periodic configurations, or by forcing them to have stabilizers that are not residually finite.

%
In the latter case (for example when the stabilizer is a Tarski monster group, see for instance \cite{ol2012geometry}), one can build SFTs with strongly periodic configurations that do not fold uniformly onto finite graphs, as shown by the following.
\begin{remark}
  $\Gamma$ is not residually finite if and only if $\Z_{\{0\}}^\Gamma$ includes an SFT containing the Cayley graph of $\Gamma$ but no finite quotient of it.
  Indeed, one can consider the SFT $\Z$ included in $\Z_{\{0\}}^\Gamma$ defined by the additional constraint that $u$ is never a cycle, where $u$ is a noncyclic path that witnesses that $\Gamma$ is not residually finite, in the sense that all finite quotients of $\Gamma$ map $u$ to $\varepsilon$.
\end{remark}


\subsection{Quotients in Cayley SFTs}
We have seen that the notions of quotients and SFTs are linked, because finite-type constraints cannot easily tell a graph from some of its quotients.
On the opposite point of view, it is noticeable that some SFTs admit a maximal support with respect to coverings.

\begin{proposition}\label{p:cayley}
	Let $\Gamma$ be a finitely presented group, $X$ a graph, and $\Z$ some SFT defined by $\Gamma$-NN constraints.
	The following are equivalent:
	\begin{enumerate}
		\item\label{i:csft} $X\in\Z$.
		\item\label{i:qgra} $X$ is covered by some colored Cayley graph $Y$ of $\Gamma$ which is in $\Z$.
	\end{enumerate}
\end{proposition}

\begin{remark}\label{r:normstab}
  In graphs of $\Z^\Gamma_\Sigma$, all stabilizers are subgroups of $\Gamma$ (because the concatenation law is coherent with that of the group).
  One has to be careful though, that quotients of Cayley graphs may not be Cayley graphs, because the stabilizer is not always a normal subgroup.
  Nevertheless, it is classical that every finite-index subgroup includes a finite-index normal subgroup (see for instance \cite[Lemma~2.1.10]{cagrp}), so that a graph of $\Z^\Gamma_\Sigma$ is strongly periodic if and only if some stabilizer subgroup is a finite-index normal subgroup of $\Gamma$.
\end{remark}
In particular, in graphs of $\Z^{\setZ^2}_\Sigma$, all stabilizers are subgroups of $\setZ^2$
, which makes many properties easier.


\begin{proposition}\label{p:dicho}
	Let $\Z$ be an SFT defined by $\setZ^2$-NN constraints.
	Then exactly one of the following occurs:
	\begin{enumerate}
		\item\label{i:aper} Every graph in $\Z$ is supported by the infinite grid, and is strongly nonperiodic.
		\item\label{i:fini} $\Z$ contains a finite graph and a strongly periodic infinite grid.
	\end{enumerate}
\end{proposition}

  It is already interesting to note that the dichotomy from Proposition~\ref{p:dicho} is nontrivial, directly from the existence of strongly aperiodic SFTs over the grid (and other ones).
  But the next subsection will show a stronger result.

\subsection{Undecidability}

\begin{theorem}\label{t:undec}
	The dichotomy from Proposition~\ref{p:dicho} is undecidable, given a list of $\setZ^2$-NN constraints.
\end{theorem}
This is a consequence of Theorem \ref{t:perfin}, allowing us to reduce the previous dichotomy to a classical periodicity problem on $\setZ^2$.
Consequently, the class of graph SFTs that contains only infinite graphs with the same support is computably unseparable from the class of graph SFTs containing both finite and infinite graphs.

This means the undecidability of any property that is implied by one of the two classes, and which implies the negation of the other one.
\begin{corollary}
	The problem whether a graph SFT contains only infinite graphs, and the problem whether a graph SFT admits only one support for all its graphs, are both undecidable (even when restricted to graph SFTs with $\setZ^2$-NN constraints). 
\end{corollary}
More precisely, the reduction directly gives that all properties separating the dichotomy are $\Sigma^0_1$-hard.
The encoding of Cayley graphs into graph SFTs also, more directly, implies the following undecidability result.

\begin{theorem}\label{t:wpb}
	There exists a finite presentation of a group $\Gamma$ such that the problem whether a given word $u\in\Pi^*$ represents a cycle from the origin back to itself in every graph of $\Z^\Gamma_{\{0\}} $ is undecidable.
\end{theorem}
It is actually $\Sigma^0_1$-complete.


\section{Conclusion}
To sum up, we provided a formalism generalizing SFTs on Cayley graphs by allowing the support graph to be determined by forbidden subgraphs. The generalization also allows the SFT to have finite configurations and we provided examples of such SFT. By defining quotient graphs, we have shown that the only way for an SFT to not have finite configurations is to only have configurations that are either nonperiodic, or with a stabilizer that is not residually finite. Similarly, we have shown that having only one support graph requires configurations to have finite stabilizers. We used both results to prove that the finite configuration problem and the support graph unicity problem are undecidable.

In the future, we wish to study determinism and the relation between causal graph dynamics and graph subshifts, in the perspective of proving structure and complexity results on causal graph dynamics.
Moreover, graph subshifts could provide a new point of view for studying group subshifts.
On the other hand, many notions and results known over group subshifts such as sofic subshifts, determinism or entropy, would be interesting to lift to graph subshifts.

We have seen that graph subshifts could generalize both classical symbolic dynamics and finite-type group geometry. They could be used to simulate a larger class of objects; let us discuss one example.
Let $\T$ be a finite set of \emph{tiles}, \ie polygons of the plane $\setR^2$.
A \emph{tiling} is then a covering of $\setR^2$ by translated copies of tiles from $\T$, the intersection of any two of them being either empty, or a vertex, or a whole edge, or the whole tile.
Now let us define subshift $\Z_\T$ with $\Sigma=\T$, $\pi$ as the set of tile edges up to translation, and the following constraints: each vertex labeled $t\in\T$ must have exactly those ports which correspond to the edges of $t$, and every simple cycle must correspond to a valid vertex in a tiling of the plane by $\T$ (\ie it involves tiles that can be put next to each other around this vertex).
The fact that the latter condition can be forced by finitely many constraints (so that $\Z_\T$ be an SFT) is equivalent to the classical \emph{finite-local-complexity} property of $\T$.
With this construction, there is a one-to-one correspondence between tilings by $\T$ of the plane and of all surfaces and graphs in $\Z_\T$.
Now if one is interested in the shift-orbit closure of a single tiling of the plane (\ie the Penrose tiling), it is an interesting question whether this corresponds to an SFT in $\Z_\T$.
Some work has been devoted to understanding for which tilings one can force locally tilings of the plane to be in the shift-orbit closure \cite{fersab,ferbed}; 
one says that the tilings admits \dfn{local rules}.
Additionnally forbidding (via finite constraints) all possible graphs which represent tilings on surfaces but not on the plane is a new question.
Our work shows that if the tiles are rectangles and if we allow some decorations (that could also be encoded locally as small hanging subgraphs), then this is possible.
It is tempting to think that this remains possible in the case of general strongly nonperiodic tilings, but the proofs would require a new formalism, because the support is no longer a Cayley graph.
The same could be done in the tridimensional space, or in the hyperbolic plane.

Let us make a final remark.
Graph theorists are used to considering families of graphs defined by {forbidden finite graph minors}.
Of course, seen as including possibly infinite graphs, these families are subshifts, but they are a priori not SFTs (not even sofic). However, even if the \emph{minor} operation involves vertices at possibly unbounded distance, it is known to have some kind of representation as local constraints \cite{kuskelohrey} (involving a nondeterministic coloring). The relation with our setting could be investigated.

{\subsection*{Acknowledgements}

We thank the referees for their careful reading of the preliminary version, and helpful remarks.
We are indebted to Nicolas Schabanel who ecouraged us to study graph subshifts and contributed to the first four definitions. This publication was made possible through the support of the ID\# 62312 grant from the John Templeton Foundation, as part of the \href{https://www.templeton.org/grant/the-quantum-information-structure-of-spacetime-qiss-second-phase}{‘The Quantum Information Structure of Spacetime’ Project (QISS) }. The opinions expressed in this publication are those of the author(s) and do not necessarily reflect the views of the John Templeton Foundation.}

\bibliography{bibliography,biblio}
\newpage\appendix
\section{Definitions}
\subsection{Topology on pointed graphs modulo}
Having a well-defined notion of disks allows us to define a topology upon ${\cal X}$, which is the natural generalization of the well-studied Cantor metric upon CA configurations or tilings \cite{Hedlund}.
\begin{definition}[Gromov-Hausdorff-Cantor metrics]\label{def:metric}
Consider the function
\begin{align*}
d:{\cal X}\times{\cal X} &\longmapsto {\mathbb R}^+\\
(X,Y)&\mapsto d(X,Y)=0\quad \textrm{if }X=Y\\	
(X,Y)&\mapsto d(X,Y)=1/2^r\quad \textrm{otherwise}
\end{align*}
where $r$ is the minimal radius such that $X^r \neq Y^r$ (\ie smallest distance from the origin to a difference between $X$ and $Y$).
\end{definition}
The function $d(.,.)$ is such that for $\eta>0$ we have (with $r=\lfloor - \log_2(\eta)\rfloor$):
$$d(X,Y)<\eta \iff X^r = Y^r.$$

\begin{remark}
  This defines an ultrametric distance, and the topology can be seen as a product of discrete topologies over the (unions of finitely many) disks of each radius.
\end{remark}

\subsection{Subshifts}
\begin{proof}[of Lemma~\ref{l:sft}]
{
~\begin{description}
		\item [$\ref{i:forbs}\Rightarrow\ref{i:forb}$] For every finite $\F$, define $M\defeq\bigcup_{(F,L)\in \F} L$ and $\F'\defeq(\X\setminus\Z)\restr{M}$. $\F'\times\{M\}$ defines the same subshift as $\F$, and $\F'$ is finite because $M$ is finite.
		\item [$\ref{i:allos}\Rightarrow\ref{i:allo}$] Similarly, for every $\A$, define $M\defeq\bigcup_{(F,L)\in \F} L$ and $\A'\defeq\Z\restr{M}$. $\A'\times\{M\}$ defines the same subshift as $\A$.
		\item [$\ref{i:forb}\Leftrightarrow\ref{i:allo}$]  The subshift $\Z$ obtained by forbidding $\F' \times\{M\}$ is the same as that obtained from allowing $\A' \times\{M\}=(\X\restr{M}\setminus \F') \times\{M\}$. Note that $\X\restr M$ is finite.
		\item [$\ref{i:forb}\Rightarrow\ref{i:forbs};$] $\ref{i:allo}\Rightarrow\ref{i:allos}$ Obvious.
\end{description}}
\end{proof}

\subsection{Groups of paths, cycles, periods}
{
  \begin{remark}\label{r:cyclgr}
    A \dfn{reduced path} is a path that does not go through the same edge twice consecutively in opposite directions, \ie $u.abba.v$ is not simple as it can be reduced to $u.v$. The language up to this reduction equivalence is a groupoid for concatenation (and inversion), \ie alike a group but its law is partially defined.\\
    The set of \dfn{(centered) cycles}, \ie reduced paths from the pointer back to it, is a subgroup of it (concatenation and inversion are always defined among them).\\
    The set of vertices is the quotient of the set of reduced paths by the following equivalence relation: $u\equiv_X v$ if and only if there exists a cycle $c$ such that $u=c.v$. Indeed, whenever they designate the same vertex $\hat u=\hat v$, we have $u=(u\ov v)v$, where $u\ov v$ is a cycle.\\
	Nevertheless, concatenation, does not induce a semigroup at the level of vertices, since we may have $\hat u=\hat v$ and yet $\widehat{wu}\ne\widehat{wv}$.
  \end{remark}
}
{
  \begin{remark}\label{r:ppergr}
   The set of reduced path-periods is a subgroup of the groupoid of reduced paths defined in Remark~\ref{r:cyclgr}. The set of cycles is a normal subgroup of the group of path-periods (since if $u$ is a path-period and $v$ is a cycle, then $\ov uv u$ is still a cycle).
  Like in Remark~\ref{r:cyclgr}, the stabilizer is the quotient of the group of reduced path-periods by the normal subgroup of cycles (since if $u$ and $v$ are paths to the same vertex, then $u=(u\ov v)v$, where $u\ov v$ is a cycle).
\end{remark}
All the statements in Remark~\ref{r:ppergr} are easy to prove, but w}e concentrate on proving that concanation and inversion (defined in Section~\ref{sec:graphs}) behave nicely over periods, which will be useful for the rest of the paper.

\begin{proposition}\label{p:group}
  Let $X$ be a graph.
  For $\hat{u},\hat{v}$ in $\stab{X}$, the concatenation operation $\hat{u}\conc\hat{v}\mapsto\widehat{u\conc v}$, and the reverse operation $\hat{u}\mapsto\widehat{\ov u}$ are well-defined, and endow $\stab X$ with a group structure.
\end{proposition}
\begin{proof}~
  \begin{itemize}
  \item Let $u\equiv_X u'$ and $v\equiv_X v'$.
    Remark~\ref{r:equiv} already gives that $uv\equiv_Xu'v$.
    Now from $X_{u'}=X$, one can deduce that $v\equiv_{X_{u'}} v'$, which means that $u'v\equiv_X u'v'$.
    We have proven that $uv\equiv_X u'v'$, so that the concatenation operation is well defined over vertices.
  \item Remark~\ref{r:equiv} gives that $u'\ov{u'}=\varepsilon=u\ov u\equiv_Xu'\ov u$.
    This means that $\ov{u'}\equiv_{X_{u'}}\ov u$, and since $X=X_{u'}$, $\ov{u'}\equiv_X\ov u$.
    We have proven that the inversion is well defined over vertices.
\item \emph{Neutral element}: the origin $\hat{\varepsilon}$ is a period, because $X_\varepsilon=X$.
\item \emph{Concatenation}: If $\hat{u},\hat{v}\in\stab{X}$, then $(X_{uv}) = (X_{u})_{v}=X_v=X$. It follows that $\widehat{u\conc v}$ is a period.
\item \emph{Inversion}: {If $\hat{u} \in \stab{X}$, then since $X = X_{u\conc\ov{u}}  = (X_u)_{\ov{u}} = X_{\ov{u}}$. It follows that $\widehat{\ov{u}}$ is a period. This period provides a right and left inverse for $\hat{u}$, since $\widehat{u\conc\ov{u}}=\hat{\varepsilon}=\widehat{\ov{u}\conc\ov{\ov{u}}}=\widehat{\ov{u}\conc u}$.}
\end{itemize}
\end{proof}

\section{Stabilizer, quotient}
\begin{proof}[of Proposition~\ref{p:quotient}]
First of all, notice that $H\conc\hat{u}=H\conc\hat{u'}$ implies that 
 there exists $\hat{h}\in H$ such that $\hat{u'}=\hat{h}\conc \hat{u}$, so that $X_{u'}=(X_{h})_{u}=X_{u}$.
\\
Let us prove that $\sigma_{\gquotient{X}{H}}$ is well-defined.
Let $\hat{u},\hat{u'}\in V(X)$ such that $H\conc \hat{u}=H\conc \hat{u'}$.
From the remark above, $X_u=X_{u'}$, and therefore $\sigma_{{X}}(\hat u)=\sigma_{{X}}(\hat u')$.\\
Now let us prove that if there is an edge $\{H\conc \hat{u}:a,H\conc \hat{v}:b\}$ in $G(\gquotient{X}{H})$, then there is an edge $\{\hat{u}:a,\widehat{h''}\conc \hat{v}:b\}$ in $G(X)$.
Indeed, by definition and the remark above, there is an edge $\{\hat{h}\conc \hat{u}:a,\hat{h'}\conc \hat{v}:b\}$ in $G(X)$ for some $\hat{h},\hat{h'}\in H$.
Since $X_u=X_{hu}$, there is also an edge $\{\hat{u}:a,\hat{\ov h}\hat{h'}\hat{v}:b\}$.
Thus $\L(X)=\L(X/H)$, and $H\conc \hat{u}$ inherits the absence of port conflict from $\hat{u}$.
The function $\varphi$ is then a covering.
\end{proof}

\begin{remark}\label{r:conjquot}
	If $H$ is a stabilizer subgroup of $X$, then for every path $u\in \L(X)$, ${\ov u}\conc H\conc u$ is a stabilizer subgroup of $X_u$.
	$\gquotient{(X_u)}{\ov{u}H\conc u}$ is well-defined and turns out to be equal to $(\gquotient{X}{H})_u$.
\end{remark}

\begin{proof}[of Proposition~\ref{p:rseparate}]
		By definition, if $u$ is a valid path in $X$, then it is also a valid path in $\gquotient{X}{H}$, and the internal state $\sigma_X(\hat u)=\sigma_{\gquotient{X}{H}}(\hat u)$ is the same.
		It remains to prove that if $v,w$ designate distinct vertices in $(X_u)\restr{M}$, then they still designate distinct vertices in $(\gquotient{X}{H}_u)\restr{M}$.
		Assume on the contrary that ${v}\equiv_{(\gquotient XH_u)\restr M}{w}$.
                By the fact that concatenation is well defined (Proposition~\ref{p:quotient}), there exists $h$ such that $\hat h\in H$ and $h\conc v\equiv_{X_u}w$, \ie $h\equiv_{X_u}w\bar v$.
                Since $w\bar v\in(X_u)\restr{M\conc\ov M}$ and $\hat h\in H$, separation gives that $w\bar v\equiv_{X_u}h\equiv_{X_u}\varepsilon$.
                This means that $w\equiv_{X_u}v$.
                This proves that the homomorphism induces an isomorphism over $(X_u)\restr M$.
                By definition, this homomorphism does not change the labeling alphabet, so that (modulo isomorphism) $(X_u)\restr M=(\gquotient XH_u)\restr M$.
                If $\X$ has defining window $M$, then by definition $X\in\X\iff\gquotient XH\in\X$.
\end{proof}

\begin{proof}[of Lemma~\ref{l:finquot}]
First, assume that $H\defeq\stab X$ is $M$-dense for some $M\subfini\Pi^*$:
For every $u\in \L(X)$, there exist $\hat{h}\in\stab X$ and $v\in\ov M$ such that $\hat{u}$ can be written $\hat{h}\conc \hat{v}$.
Now for every $\hat{h'}\in H$, $\hat{h'}\conc \hat{u}=\hat{h'}\conc \hat{h}\conc \hat{v}\in H\conc \hat{v}$, so $H\conc \hat{u}\subseteq H\conc \hat{v}$ and $\hat{h'}\conc \hat{v}=\hat{h'}\conc \hat{\ov h}\conc \hat{h}\conc \hat{v}=\hat{h'}\conc \hat{\ov h}\hat{u}\in H\conc \hat{u}$, so $H\conc \hat{v}\subseteq H\conc \hat{u}$. Thus $H\conc \hat{u}=H\conc \hat{v}$.
Hence, the set $\sett{H\conc \hat{u}}{\hat{u}\in V(X)}=\sett{H\conc \hat{v}}{v\in M\cap \L(X)}$ of vertices of $\gquotient{X}H$ is finite. 
We conclude by Proposition~\ref{p:quotient}.\\
{Conversely, if $X$ admits a finite quotient $\gquotient XH$, there exists $M$ such that $\sett{H\conc \hat{u}}{{u}\in \L(X)}=\sett{H\conc \hat{v}}{v\in M\cap \L(X)}$.
	For all $\hat{u}\in V(X)$, since $\hat{u}\in H\conc \hat{u}$, we get that $\hat{u}\in H\conc \hat{v}$ for some $v\in M$. Therefore, $\hat{u}=\hat{h}\conc \hat{v}$ with $\hat{h}\in \stab X$ and $v\in M$}, so that $H$ is $\ov M$-dense.
    \end{proof}

\begin{proof}[of Theorem~\ref{t:perfin}]
The left-to-right implication directly comes from the fact that finite graphs are strongly periodic and finite groups are residually finite.

For the right-to-left direction, suppose that $\Y$ has defining window $M$ and admits a strongly periodic graph $X$ such that $\stab{X}$ is residually finite.
Since $X$ is strongly periodic, $\bigcup_{\hat u\in V(X)}V((X_u)\restr{M\ov M})$ is a finite union of finite subgraphs.
Since $\stab{X}$ is residually finite, there exists a subgroup $H\le\stab{X}$ of finite index which does not intersect $\bigcup_{\hat u\in V(X)}V((X_u)\restr{M\conc\ov M})\setminus\{\varepsilon\}\subseteq\Pi^{M\conc\ov M}$.
As chosen, $H$ is $M$-separated, so that we can use Proposition~\ref{p:rseparate}: for all $\hat{u}\in V(X)$ we have that $(\gquotient{X}{H}_u)\restr M=(X_u)\restr M$. Given that $\Y$ has defining window $M$, we get that $\gquotient{X}{H}\in\Y$, and $\gquotient XH$ is finite by Lemma~\ref{l:finquot}.
\end{proof}

\begin{proof}[of Proposition~\ref{p:cayley}]~\begin{description}
		\item[\ref{i:csft}$\Rightarrow$\ref{i:qgra}]
		Suppose that the support of $X$ is in $\Z$.
	Let us describe a color for each vertex $\hat u$ of a Cayley graph $Y$ of $\Gamma$.
	We note that paths designating vertex $\hat u$ in $Y$ all designate the same vertex in $X$: indeed, if $\hat v$ also designates $\hat u$ in $Y$, then it can be obtained from $\hat u$ by iteratively applying the reduction rules from the presentation.
	Then thanks to the third condition in the definition of $\Z^\Gamma_\Sigma$, it also designates the same vertex as $\hat u$ in $X$.
	Hence one can simply define $\phi(\hat u)=\hat u\in V(X)$ for every $\hat u\in V(Y)$, and $\sigma_Y(\hat u)\defeq\sigma_X(\hat u)$.
        This clearly satisfies the conditions of Definition~\ref{d:morph}.
	\item[\ref{i:qgra}$\Rightarrow$\ref{i:csft}]
          Conversely, consider a graph $X$ covered by some Cayley graph $Y$ of $\Gamma$ which is in $\Z$.
          Consider a cycle constraint of $\Z$, that is: every path labeled by some $u$ (in the group presentation of $\Gamma$) must be a cycle.
          This constraint is satisfied in $X$ because loops are preserved by homomorphism.
          Now consider an edge constraint of $\Z$. These are also preserved by homomorphism, by definition.
          We have proven that $X$ satisfies all constraints defining $\Z$.
	\end{description}
\end{proof}

\begin{proof}[of Proposition~\ref{p:dicho}]
  Assume that Condition~\ref{i:aper} is not satisfied.
	A classical pumping argument from multidimensional symbolic dynamics (see for instance \cite[Lemma~1.16]{ballier}) shows that $\Z$ then contains a strongly periodic configuration. 
	In that case, its stabilizer, as a subgroup of $\setZ^2$, is residually finite (see for example \cite[Prop.2.2.1]{cagrp}).
	Theorem~\ref{t:perfin} then implies that $\Z$ contains a finite graph.
	
	On the other hand, it is clear that both conditions cannot be satisfied simultaneously, so that they indeed form a dichotomy.
\end{proof}

\subsection{Nearest-neighbor constraints on directed regular SFT}
{
In contrast with $\Z^{\setZ^2}_\Sigma$, $\Z^{\setF_2}_\Sigma$ is much less rigid in the sense of the following statement, because it can be related to tilings over square-tiled surfaces, as studied in \cite{gambaudo,downhier}, and formalized in \cite[Subsection~1.3.3]{jeandel-vanier}.
\begin{proposition}
  Let $\Z$ be an SFT defined by $\setF_2$-NN constraints.
  Then the following are equivalent:
  \begin{enumerate}
  \item\label{i:nonv} $\Z$ is nonempty.
  \item\label{i:wper} $\Z$ contains a weakly periodic Cayley graph of $\setF_2$.
  \item\label{i:smallcg} $\Z$ contains a graph whose support is not the Cayley graph of $\setF_2$.
  \end{enumerate}
  Besides, the following are equivalent:
  \begin{enumerate}[i.]
  \item\label{i:finite} $\Z$ contains a finite graph.
  \item\label{i:finicg} $\Z$ contains a finite Cayley graph.
  \item\label{i:strper} $\Z$ contains a strongly periodic Cayley graph of $\setF_2$.
  \item\label{i:amen} $\Z$ contains an amenable Cayley graph.
  \end{enumerate}
  {Nevertheless, there exist nonempty SFTs containing only infinite graphs and defined by $\setF_2$-NN constraints.}
\end{proposition}
The second equivalence is very close to the main result of \cite{addendum}, about how tilesets tile Cayley graphs. They also show the equivalence with containing a \dfn{cylindric} graph, \ie with having an infinite cyclic dense stabilizer.
\begin{proof}~\begin{description}
  \item[\ref{i:nonv}$\Rightarrow$\ref{i:wper}]If $\Z$ is nonempty, then by Proposition~\ref{p:cayley}, it contains a Cayley graph of $\setF_2$.
    It is known that every SFT over $\setF_2$ contains a weakly periodic configuration (see for instance \cite[Theorem~2.2]{piantadosi}).
  \item[\ref{i:wper}$\Rightarrow$\ref{i:smallcg}]Let $X$ be a weakly periodic Cayley graph of $\setF_2$ and $\hat u$ be a nontrivial period.
    Then the infinite cyclic group $\left<\hat u\right>$ is a residually finite stabilizer subgroup: if $M$ is a defining window for $\Z$, by the same construction as for Theorem~\ref{t:perfin}, there exists an infinite (cyclic) subgroup $H<\left<\hat u\right>$ which is $M$-separated.
    By Proposition~\ref{p:rseparate}, $X/H$ belongs to $\Z$.
  \item[\ref{i:smallcg}$\Rightarrow$\ref{i:nonv}]This is trivial.
  \item[\ref{i:finicg}$\Rightarrow$\ref{i:finite}]This is trivial.
  \item[\ref{i:strper}$\Rightarrow$\ref{i:finicg}]Let $X$ be some strongly periodic Cayley graph of $\setF_2$.
    By the proof of Theorem~\ref{t:perfin}, there exists a finite-index subgroup $H$ of $\stab(X)$ which is $M$-separated, where $M$ is a defining window for $\Z$.
    Thanks to Remark~\ref{r:normstab}, $H$ includes a finite-index normal subgroup $H'$ of $\setF_2$:
    $H'$ is still $M$-separated, and $X/H$ is a finite Cayley graph (concatenation is coherent with the quotient rule).
    By Proposition~\ref{p:quotient}, $X/H$ is in $\Z$.
  \item[\ref{i:finite}$\Rightarrow$\ref{i:strper}]If $\Z$ contains a finite graph $X$, then from Proposition~\ref{p:cayley}, it also contains a colored Cayley cover $Y$ of $X$.
    All preimages of $\varepsilon$ by the homomorphism are period in $Y$ (because they are periods in $X$), and their set is $M$-dense, where $M$ is a finite language representing all possible paths in $X$.
  \item[\ref{i:finicg}$\Rightarrow$\ref{i:amen}]This comes from the fact that finite groups are amenable.
  \item[\ref{i:amen}$\Rightarrow$\ref{i:finite}]This is remarked in \cite{downhier}, based on \cite{gambaudo}.
    {  \item The last statement is also inherited from symbolic dynamics over groups \cite[Example~3]{piantadosi}, thanks to the above equivalence.}
  \end{description}\end{proof}
}

\begin{proof}[of Theorem~\ref{t:undec}]
	Every set $\F$ of $\setZ^2$-NN constraints defines both a (graph) SFT $\Z$, and a classical SFT $\Z'$ over the infinite grid (which is included in $\Z$).
	If $\Z'$ is strongly aperiodic, then $\Z$ is in Class~\ref{i:aper}, whereas if it contains a weakly periodic configuration, then $\Z$ is in Class~\ref{i:fini}.
        Hence if there were an algorithm to separate the two classes, then we could use it to decide wheterh $\Z'$ is strongly aperiodic or not, which is known to be undecidable \cite{gurkor72}.
%
\end{proof}

\begin{proof}[of Theorem~\ref{t:wpb}]
  From Proposition~\ref{p:cayley}, $u$ represents a cycle from the origin back to itself in every graph of $\Z^\Gamma$ if and only if it represents the identity element in $\Gamma$, which means that it is in the language of the \emph{word problem}.
  If we could decide this property, then we could decide the word problem, which is known to be impossible in some finitely presented group $\Gamma$ \cite{wordpb}.
\end{proof}
\end{document}